\documentclass[reprint,amsmath,amssymb,aip]{revtex4-1}
\makeatletter
\let\@fnsymbol\@fnsymbol@latex
\@booleanfalse\altaffilletter@sw
\makeatother
\usepackage{graphicx}% Include figure files
\usepackage{dcolumn}% Align table columns on decimal point
\usepackage{bm}% bold math
\usepackage[usenames,dvipsnames]{color}

\def \be {\begin{equation}}
\def \ee {\end{equation}}
\def \ben {\begin{eqnarray}}
\def \een {\end{eqnarray}}
\def \bi {\begin{itemize}}
\def \ei {\end{itemize}}

\usepackage{multirow}
\usepackage{ctable}
\begin{document}
\bibliographystyle{prsty}

%\preprint{APS/123-QED}

\title{A simple generalization of the energy gap law for nonradiative processes}

\author{Seogjoo J. Jang}
\email[ ]{seogjoo.jang@qc.cuny.edu}

\affiliation{Department of Chemistry and Biochemistry, Queens College, City University of New York, 65-30 Kissena Boulevard, Queens, New York 11367\footnote{mailing address}  \& PhD programs in Chemistry and Physics, and Initiative for the Theoretical Sciences, Graduate Center, City University of New York, 365 Fifth Avenue, New York, NY 10016}

\date{Published in {\it the Journal of Chemical Physics} {\bf 155}, 164106  (2021)}

\begin{abstract}
For more than 50 years, an elegant energy gap (EG) law developed by Englman and Jortner [Mol. Phys. {\bf 18}, 145 (1970)] has served as a key theory to understand and model nearly exponential dependence of nonradiative transition rates on the difference of energy between the initial and final states.  This work revisits the theory, clarifies key assumptions involved in the rate expression, and provides a generalization for the cases where the effects of temperature dependence and low frequency modes cannot be ignored. For a specific example where the low frequency vibrational and/or solvation responses can be modeled as an Ohmic spectral density, a simple generalization of the EG law is provided.  Test calculations demonstrate that this generalized EG law brings significant improvement over the original EG law.   Both the original and generalized EG laws are also compared with stationary phase approximations developed for electron transfer theory, which suggests the possibility of a simple  interpolation formula valid for any value of EG.

\end{abstract}

\maketitle
\section{Introduction}
In 1970, Englman and Jortner (EJ)\cite{englman-mp18} and also Fischer\cite{fischer-jcp53} derived an elegant energy gap (EG) law\cite{englman-mp18,fischer-jcp53,freed-acr11} explaining nearly exponential dependence of the nonradiative decay rate on the transition energy.  The theory and model was then extended and improved further,\cite{gelbart-jcp52,freed-jcp52,fong-jcp56,nitzan-jcp63,freed-acr11,lin-jcp58,kober-jpc90} but the simple form\cite{englman-mp18} of the first EG rate law and model continued being used widely to account for quantum yields of excited states and  to estimate energy levels of dark states involved in many radiationless transitions.\cite{martin-cpl35,caspar-cpl91,caspar-jpc87,wilson-jacs123,chynwat-cp194,asahi-jpc97}  With emerging demands for better assessment and control of quantum yields and luminescence properties of molecules and materials, for example, for the improvement of solar energy conversion,\cite{demchenko-cr117} light emitting devices,\cite{pan-aom2,wei-np14,zampetti-afm29} and imaging,\cite{hong-nbe1} there is now new interest for more accurate spectroscopic assessment\cite{friedman-chem} and detailed theoretical/computational analyses of the EG law.

There were examples\cite{kober-jpc90} for which the EG law, with the inclusion of Franck-Condon factors, have provided unambiguous and quantitative explanation of experimental results.  However, more often than not, exact details of how and why the EG law works despite other complicating factors have remained poorly understood.  In some cases, different explanations\cite{schlag-arpc22,heller-jcp79} offering alternative scenarios were suggested as well.  Thus, it is true that there have been rather insufficient theoretical analyses or computational studies of the EG law despite its long history.  This motivates more careful examination of theoretical assumptions behind the EG law and consideration of its extension for more general cases.  

The objective of this work is to revisit the EG law employing a generic spin-boson type model Hamiltonian and then to examine its corrections when the conditions for the original EG law\cite{englman-mp18} are not fully satisfied.   Section II provides a review of the EG law.  Section III presents a simple generalization of the EG law for the cases with additional Ohmic environments.  Section IV provides model calculations illustrating the utility of new expressions derived here.  Section V concludes the paper.  

\section{Energy Gap law: Stationary Phase Approximation for High Frequency Single Vibrational Mode}
The EG law\cite{englman-mp18} is an application of the stationary phase approximation for the Fermi Golden Rule (FGR) rate expression.  There are two assumptions underlying the EG law.  One is that the electronic transition of interest is  weakly coupled to molecular vibrations and environmental degrees of freedom. The other is that the vibronic transition due to the highest frequency vibrational modes serves as the main route for the quantum transition mechanism.     To clarify this, let us consider the following minimal Hamiltonian:
\ben 
H&=&E_1|1\rangle\langle 1| +E_2|2\rangle\langle 2|+J(|1\rangle\langle 2|+|2\rangle\langle 1|) \nonumber \\
&+&B_1|1\rangle\langle 1|+B_2|2\rangle \langle 2|+H_b , \label{eq:hamiltonian}
\een
where $|1\rangle$ is the initial excited state with energy $E_1$ and $|2\rangle$ is the final state with energy $E_2$ to which transition from $|1\rangle$ occurs due to the electronic coupling $J$. 
$H_b$ is the bath Hamiltonian representing the vibrational modes and environmental responses altogether, and $B_1$ ($B_2$) is the coupling term of the bath to state $|1\rangle$ ($|2 \rangle$). 
EJ considered\cite{englman-mp18}  the cases where the electronic coupling $J$ originates from either nonadiabatic or spin-orbit couplings, but the only assumption needed for deriving the EG law is that     
$J$ is a constant small enough to validate the application of FGR. For the Hamiltonian of Eq. (\ref{eq:hamiltonian}), this FGR rate can be expressed as\cite{jang-exciton}  
\ben
k_{FG}&=&\frac{2J^2}{\hbar^2} {\rm Re} \Big [ \int_0^\infty dt\ e^{i \Delta E t/\hbar} \nonumber \\
&&\times Tr_b \left \{ e^{i H_b+B_1)t/\hbar} e^{-i(H_b+B_2)t/\hbar} \rho_{b1} \right\} \Big ] , \label{eq:kfg-1}
\een
where $\rho_{b1}=e^{-(H_b+B_1)/(k_BT)}/Tr_b\{e^{-(H_b+B_1)/(k_BT)}\}$ and 
\be
\Delta E=E_1-E_2 .
\ee
Let us consider the simple case that $H_b$ consists of harmonic oscillators and $B_1$ and $B_2$ are linear in the displacements of the bath oscillators.  Namely,  $H_b= \sum_n \hbar\omega_n (b_n^\dagger b_n+\frac{1}{2})$,  $B_1= \sum_n \hbar\omega_n g_{n1} (b_n+b_n^\dagger)$,  and $B_2= \sum_n \hbar\omega_n g_{n2} (b_n+b_n^\dagger)$, where $\omega_n$ is the angular frequency and $b_n$ ($b_n^\dagger$) is the lowering (raising) operator of the $n$th mode, and $g_{n1}$ ($g_{n2}$) is the dimensionless coupling strength\cite{jang-exciton,jang-rmp90} (displacement) of the $n$th mode to the state $1$ ($2$).  For this case, Eq. (\ref{eq:kfg-1}) can be evaluated exactly\cite{jang-exciton} and becomes
\ben
&&k_{FG}=\frac{J^2}{\hbar^2}\int_{-\infty}^\infty dt\ \exp \Bigg [\frac{i}{\hbar}\Delta \bar E  t-\sum_n \delta g_n^2   \nonumber \\
&& \times \Big \{ \coth \left (\frac{\hbar\omega_n}{2k_BT} \right) \left (1-\cos (\omega_n t)\right) +i\sin (\omega_n t) \Big\} \Bigg] , \label{eq:kfg-2}
\een
where $\delta g_n=g_{n1}-g_{n2}$ and  
\be
\Delta \bar E = \Delta E-\sum_n \hbar\omega_n (g_{n1}^2-g_{n2}^2) .
\ee
The term for lifetime decay is omitted in Eq. (\ref{eq:kfg-2}), which however can be added when necessary, either by assuming an imaginary component in $E_1$ or explicitly introducing the background radiation via quantum electrodynamics formulation.\cite{craig}   

For the evaluation of Eq. (\ref{eq:kfg-2}) using the stationary phase approximation, it is useful to rewrite Eq. (\ref{eq:kfg-2}) as follows:
\be
k_{FG}=\frac{J^2}{\hbar^2} e^{-f(0)}\int_{-\infty}^\infty dt\ e^{f(t)} , \label{eq:kfg-3}
\ee
where
\be
 f(t)=\frac{i}{\hbar}\Delta \bar E t  +\sum_n \delta g_n^2 \frac{\cosh\left (\omega_n(\frac{\hbar}{2k_BT}-i t) \right )}{\sinh \left (\frac{\hbar\omega_n}{2k_BT}\right ) } .\label{eq:ft}
\ee
 Then, deforming the integration for $t$ in Eq. (\ref{eq:kfg-3}) to the complex domain and making the stationary phase approximation along the steepest descent path around a stationary point,\cite{morse-feshbach-1} Eq. (\ref{eq:kfg-3}) can be approximated as 
 \be
k_{SP}= \frac{J^2}{\hbar^2} \left (\frac{2\pi}{-f''(t_s)} \right)^{1/2} e^{-f(0)+f(t_s)} , \label{eq:kfg-3-1}
\ee
where  $t_s$ is the stationary phase value for which $f'(t_s)=0$ and $f''(t_s)$ is the second derivative at that point.  Note that the above expression is valid for any $t_s$ in complex domain.   An implicit equation for $t_s$ can be found easily by taking the derivative of Eq. (\ref{eq:ft}) directly, which results in
\be
\frac{\Delta \bar E}{\hbar}  = \sum_n \delta g_n^2 \omega_n \frac{\sinh\left (\omega_n(\frac{\hbar}{2k_BT}-i t_s) \right )}{\sinh \left ( \frac{\hbar\omega_n}{2k_BT} \right) } . \label{eq:sta-soln}
\ee
On the other hand, the second derivative of $f(t)$ is  
\ben
 f''(t_s)&=&-\sum_n \delta g_n^2 \omega_n^2 \frac{\cosh\left (\omega_n(\frac{\hbar}{2k_BT}-it_s) \right )}{\sinh (\frac{\hbar\omega_n}{2k_BT}) } \nonumber \\
 &\equiv& -D(t_s) ,\label{eq:ft-2nd}
\een
where the second equality serves as the definition of $D(t_s)$.   Employing this expression and Eq. (\ref{eq:ft}) for $t_s$ in Eq. (\ref{eq:kfg-3-1}), we obtain the following stationary phase approximation for the rate:
\ben
k_{SP}&= &\frac{J^2 }{\hbar^2}\left (\frac{2\pi}{D(t_s)}\right)^{1/2} e^{-f(0)}\exp \Bigg [\frac{i}{\hbar}\Delta \bar E t_s   \nonumber \\
&&+\sum_n \delta g_n^2 \frac{\cosh\left (\omega_n(\frac{\hbar}{2k_BT}-i t_s) \right )}{\sinh \left (\frac{\hbar\omega_n}{2k_BT}\right ) } \Bigg ].  \label{eq:kfg-4}
\een
An expression equivalent to this has been used as the starting point for many theoretical improvements\cite{gelbart-jcp52,freed-jcp52,fong-jcp56,nitzan-jcp63,freed-acr11,lin-jcp58,kober-jpc90} of the EG law and also for deriving closed form expressions for electron transfer (ET) rates in the quantum limit.\cite{jang-jpcb110} 
 
A typical situation where the EG law applies is when $\Delta \bar E$ is much larger than both the thermal energy and the reorganization energy of the vibrational modes coupled to the electronic transition.  In this case, the solution for Eq. (\ref{eq:sta-soln}) exists only for sufficiently large imaginary value of $t_s$ with the major contribution coming from the highest frequency vibrational modes.   To clarify this point and also to understand the relationship between the EG law and other stationary phase approximations, let us introduce the following standard form of the bath spectral density:
\be
{\mathcal J}(\omega)=\pi \hbar \sum_n \delta g_n^2 \delta (\omega-\omega_n) \omega_n^2 . \label{eq:spd-def} 
\ee
In addition, let us introduce an imaginary time $\tau_s$ such that
\be
\tau_s=-it_s , 
\ee 
for which, the implicit equation for $t_s$, Eq. (\ref{eq:sta-soln}), with the use of the bath spectral density ${\mathcal J}(\omega)$, becomes 
\ben
\frac{\Delta \bar E}{\hbar}  &=&\frac{1}{\pi\hbar} \int_0^\infty d\omega \frac{ {\mathcal J}(\omega)}{\omega}  \frac{\sinh\left (\omega (\frac{\hbar}{2k_BT}+\tau_s) \right )}{\sinh \left ( \frac{\hbar\omega}{2k_BT} \right) } . \label{eq:sta-soln-tau}
\een
Then, the stationary phase approximation for the FGR rate, Eq. (\ref{eq:kfg-4}), can be expressed in terms of $\tau_s$ as follows:
\be
k_{SP}= \frac{J^2 }{\hbar^2}\left (\frac{2\pi}{D(i\tau_s)}\right)^{1/2} \exp \Bigg [-\frac{1}{\hbar}\Delta \bar E \tau_s +G(\tau_s)\Bigg] , \label{eq:kfg-5} 
\ee
where
\ben
&&D(i\tau_s)= \frac{1}{\pi\hbar} \int_0^\infty d\omega {\mathcal J} (\omega)  \nonumber \\
&&\hspace{.1in} \times \left (\coth \left (\frac{\hbar\omega}{2k_BT}\right) \cosh (\omega \tau_s ) +\sinh  (\omega\tau_s)  \right )  ,  \label{eq:d-tau}\\
&&G(\tau_s)= \frac{1}{\pi\hbar} \int_0^\infty d\omega \frac{{\mathcal J}(\omega)}{\omega^2} \nonumber \\
&&\hspace{.1in}\times \left (\coth \left (\frac{\hbar\omega}{2k_BT}\right) \left ( \cosh (\omega \tau_s ) -1\right) +\sinh  (\omega\tau_s)  \right) .  \nonumber \\ \label{eq:g-tau}
\een

Stationary phase approximations developed for the ET theory can be converted to the present case by using the fact that $\Delta \bar E$ is equal to $-\Delta G$, where $\Delta G$ is the reaction free energy of the ET reaction.  Thus the three known stationary phase approximations for ET with closed form expressions \cite{hopfield-pnas71,vanduyne-cp5,siders-jacs103,lang-cp244,jang-jpcb110}correspond to the cases where $\Delta \bar E=0$ or $\pm\lambda$, where $\lambda$ is the reorganization energy defined as 
\be
\lambda\equiv\frac{1}{\pi}\int_0^{\infty} d\omega \frac{{\mathcal J}(\omega)}{\omega}  .
\ee 
Similarly, an interpolation formula developed by Jang and Newton,\cite{jang-jpcb110} which connects the three stationary phase approximations for the ET rate can also be converted for the present case.  Later, comparison of the well-known semiclassical approximation,\cite{hopfield-pnas71,jang-jpcb110} which is known to work well in the strong coupling limit and corresponds to the stationary phase approximation for $\Delta \bar E =\lambda$, and the interpolation formula\cite{jang-jpcb110} with the EG law expression will also be provided. 

Now, in order to clarify more detailed features of the EG law, let us assume that the spectral density consists of two components as follows:
\be
{\mathcal J}(\omega)=\pi\hbar \delta \bar g_h^2 \omega_h^2 \delta (\omega-\omega_h)+{\mathcal J}_l(\omega) ,
\ee  
where $\omega_h$ is the highest vibrational frequency among all the modes with $\delta \bar g_h^2 \approx \sum_{n, \omega_n\approx \omega_h }  \delta g_n^2 $, and  
${\mathcal J}_l(\omega)$ represents all the vibrational modes and environmental responses with its major contributions coming from low frequencies that satisfy $\omega < \omega_l$, where $\omega_l < \omega_h$.  Namely, $\omega_l$ is the range of frequencies characterizing ${\mathcal J}_l(\omega)$.   For this case,  the reorganization energy $\lambda$ of the total bath spectral density ${\mathcal J}(\omega)$ can be divided into two terms as follows:
\be
\lambda=\lambda_h+\lambda_l ,
\ee
where
\ben
\lambda_h&=&\hbar \omega_h \delta \bar g_h^2 , \label{eq:lambda-m}\\
\lambda_l&=& \frac{1}{\pi}\int_0^{\infty} d\omega \frac{{\mathcal J}_l(\omega)}{\omega}  . \label{eq:lambda-s}
\een
Then,  Eq. (\ref{eq:sta-soln-tau}) can be expressed as follows:
\ben
&&\Delta \bar E  = \lambda_h \Bigg \{ \exp (\omega_h\tau_s) \nonumber \\
&&\hspace{.2in}+ \left (\coth \left (\frac{\hbar \omega_h}{2k_BT}\right) -1\right) \sinh (\omega_h\tau_s) \Bigg \} \nonumber \\ 
&&+\frac{1}{\pi}\int_0^{\infty} d\omega \frac{{\mathcal J}_l(\omega)}{\omega} \Bigg \{ \exp (\omega\tau_s) \nonumber \\
&&\hspace{.2in}+ \left (\coth \left (\frac{\hbar \omega}{2k_BT}\right) -1\right) \sinh (\omega\tau_s) \Bigg \} 
, \label{eq:sta-soln-1}
\een
where the following identity has been used: $\sinh(x+y)/\sinh(x)=e^y+(\coth(x)-1)\sinh(y)$. 

In the weak coupling limit where $\Delta \bar E>>\lambda_h,\lambda_l$, the solution of Eq. (\ref{eq:sta-soln-1}) exists only for a large enough value of $\tau_s$.   
Given that the values of $\lambda_h/\lambda_l$ and $\omega_h/\omega_l$ are sufficiently large,  it is expected that the major contribution on the right hand side comes from the first term involving the $\omega_h$ frequency modes.  Assuming that $\hbar \omega_h >> k_BT$, this term can be further approximated as $\lambda_h \exp(\omega_h \tau_s)$.   Thus, under these conditions, 
the solution for Eq. (\ref{eq:sta-soln-1}) can be approximated as 
\be
\tau_s^0=-it_s^0 =\frac{1}{\omega_h} \ln \left ( \frac{\Delta \bar E}{\lambda_h} \right) . \label{eq:tau_s-0}
\ee 
Inserting this expression into Eqs. (\ref{eq:kfg-5})-(\ref{eq:g-tau}), we obtain  
 \ben
k_{EG}&\approx& \frac{ J^2 }{\hbar^2}\left (\frac{2\pi}{D(i\tau_s^0)}\right)^{1/2} \nonumber \\
&&\hspace{.1in}\times \exp \Bigg [-\frac{\Delta \bar E}{\hbar\omega_h} \ln \left (\frac{\Delta \bar E}{\lambda_h} \right ) +G(\tau_s^0) \Bigg] ,  \label{eq:kfg-eg}
\een
where
\ben
&&D(i\tau_s^0)= \delta \bar g_h^2 \omega_h^2 \Bigg \{\exp(\omega_h\tau_s^0)\nonumber \\
&&\hspace{.8in}+ \left (\coth \left (\frac{\hbar\omega_h}{2k_BT}\right) -1\right )\cosh (\omega_h \tau_s^0 )   \Bigg \}  \nonumber \\
&&\hspace{.4in}+\frac{1}{\pi\hbar} \int_0^\infty d\omega {\mathcal J}_l (\omega)  \Bigg \{\coth \left (\frac{\hbar\omega}{2k_BT}\right) \cosh (\omega \tau_s^0 ) \nonumber \\
&&\hspace{1.5in}+\sinh  (\omega\tau_s^0)  \Bigg \}  ,  \label{eq:d-tau}
\een
\ben
&&G(\tau_s^0)=  \delta \bar g_h^2 \Bigg\{\exp(\omega_h\tau_s^0) -1\nonumber \\
&&\hspace{.4in}+\left (\coth \left (\frac{\hbar\omega}{2k_BT}\right) -1\right )\left ( \cosh (\omega \tau_s^0 ) -1\right)   \Bigg\} \nonumber \\
&&+\frac{1}{\pi\hbar} \int_0^\infty d\omega \frac{{\mathcal J}_l(\omega)}{\omega^2} \Bigg \{\coth \left (\frac{\hbar\omega}{2k_BT}\right) \left ( \cosh (\omega \tau_s^0 ) -1\right)\nonumber \\
&&\hspace{1.2 in} +\sinh  (\omega\tau_s^0)  \Bigg \} . \label{eq:g-tau}
\een

Equation (\ref{eq:kfg-eg}) is as yet different from the well-known form of the EG law rate expression.\cite{englman-mp18}     However, it is possible to show that Eq. (\ref{eq:kfg-eg}) reduces to the latter if we approximate that $\coth(\hbar\omega_h/(2k_BT))\approx 1$ and entirely neglect the contribution from ${\mathcal J}_l(\omega)$.   Thus, with these approximations, it is straightforward to make the following simplifications: 
 \ben
&& G(\tau_s^0) \approx \delta \tilde g_h^2 \Bigg \{\exp \left( \omega_h\tau_s^0 \right ) - 1 \Bigg \} =\frac{\Delta \bar E -\lambda_h}{\hbar\omega_h} , \\
&&D(i\tau_s^0)\approx \delta \tilde g_h^2 \omega_h^2 \exp \left \{ \omega_h\tau_s^0\right\} =\omega_h\frac{\Delta \bar E}{\hbar}  ,
\een
where the definitions of $\lambda_h$, Eq. (\ref{eq:lambda-m}), and $\tau_s^0$, Eq. (\ref{eq:tau_s-0}), have been used. 
Thus, with the above approximations, Eq. (\ref{eq:kfg-eg}) reduces to the following EG rate expression:\cite{englman-mp18} 
 \ben
k_{EG}&=& \frac{J^2 }{\hbar}\left (\frac{2\pi}{\hbar\omega_h \Delta \bar E} \right)^{1/2}\exp \Bigg [-\frac{\lambda_h}{\hbar\omega_h} \nonumber \\
&&  -\frac{\Delta \bar E}{\hbar\omega_h} \left \{\ln \left (\frac{\Delta \bar E}{\lambda_h}\right )-1\right\} \Bigg ]   . \label{eq:keg}
\een
For the case where the state $1$ is an excited state and $2$ is the ground state, $\Delta \bar E$ corresponds to the vertical absorption energy and $2\lambda_h +2\lambda_l=E_{st}$, which is the Stokes shift of the the radiative emission peak from the absorption peak maximum. Thus, given that Eq. (\ref{eq:keg}) serves as a sole representation of the nonradiative decay of the excited state, fitting the experimental rate with the above expression can provide an information on 
\be
\gamma \equiv\frac{2\lambda_h}{E_{st}}= \frac{\lambda_h}{\lambda_l+\lambda_h} =\frac{\hbar\omega_h \delta \bar h_h^2}{\lambda_l+\hbar\omega_h \delta \bar h_h^2} , \label{eq:gamma}
\ee 
as has been indicated in a recent work.\cite{friedman-chem}

Equation (\ref{eq:keg}) is independent of temperature, a natural outcome of a process dominated by tunneling through single high frequency vibrational modes in their ground vibrational states. Actual rates in real systems may be temperature dependent due to various conditions that are not consistent with the assumptions of the EG law.  In this sense, experimental investigation of the temperature dependence of the radiationless transition rate will be important for confirming that the observed behavior is indeed consistent with the conditions assumed  for the EG law.\cite{englman-mp18}  A practical issue that makes such temperature dependence study complicated is the fact that $\Delta \bar E$ is dependent on temperature, which however can be determined through independent spectroscopic measurements.

\section{Generalization of the EG law}
\subsection{General case}
Let alone the stationary phase approximation, which is known to be fairly accurate even for intermediate coupling regime, the validity of Eq. (\ref{eq:keg}) requires the following conditions: (i) $\Delta \bar E >> \lambda_h$; (ii) $\coth(\hbar\omega_h/(2k_BT))= 1$; (iii) $\lambda_l=0$, for which $\tau_s^0$ becomes the exact solution for Eq. (\ref{eq:sta-soln-1}).  What if all three of these conditions are not satisfied?  It is still possible for the rate to show nearly exponential dependence.   If this is the case, the EG law, Eq. (\ref{eq:keg}), cannot serve as a quantitatively reliable theory even when it appears to explain experimental trends qualitatively.   
A simple extension of Eq. (\ref{eq:keg}) for such cases is instead to use Eq. (\ref{eq:kfg-eg}), which retains all the Franck-Condon contributions of other modes and finite temperature effects. However, this expression keeps $\tau_s^0$ as the solution of the stationary phase condition.   Thus, if the solution for  Eq. (\ref{eq:sta-soln-tau}) deviates significantly from $\tau_s^0$, it will not be accurate enough.  Furthermore, it is not even clear whether Eq. (\ref{eq:kfg-eg}) is well defined for all values of $\tau_s^0$ and for any kind of ${\mathcal J}_l (\omega)$.   Therefore, it is important to find a solution for $\tau_s$ that is valid for more general conditions.  

Freed and Jortner\cite{freed-jcp52} showed that it is possible to find a general solution for $\tau_s$ by solving an algebraic equation, which can be a good starting point but appears to be difficult to work with for a general bath spectral density.   Alternatively, solving Eq. (\ref{eq:sta-soln-1}) numerically is not a difficult task with current computational capability and may be the best practical solution for general situation.  As yet, it would be useful to have a closed form expression for quick theoretical understanding and interpretation of experimental data.  To this end, it is possible to find the next order correction for the solution of Eq. (\ref{eq:sta-soln-tau}) under the condition that $\tau_s$ deviates slightly from $\tau_s^0$.  Detailed expressions for this case are provided in Appendix A. 

\subsection{Ohmic spectral density with exponential cut-off for low frequency modes}
The present subsection provides a practical approximation that can be used for a wider range of $ \Delta \bar{E}$ when the low frequency spectral density can be expressed as an Ohmic spectral density with exponential cutoff as follows:
\be
{\mathcal J}_l(\omega)=\pi\hbar \eta_l \omega e^{-\omega/\omega_l} ,
\ee     
where $\omega_l <\omega_h$.   For this, 
\be
\lambda_l=\eta_l \hbar \omega_l .
\ee
Since the above Ohmic bath spectral density is commonly used for representing the effects of solvent and a collection of low frequency modes mixed with environmental degrees of freedom, a general rate expression in this case is expected to have broad applications.  

For the above Ohmic spectral density representing the low frequency modes, Eq. (\ref{eq:sta-soln-1}) becomes 
\ben
&&\Delta \bar{ E}  = \lambda_h \Bigg \{ \exp (\omega_h\tau_s) \nonumber \\
&&\hspace{.5in}+ \left (\coth \left (\frac{\hbar \omega_h}{2k_BT}\right) -1\right) \sinh (\omega_h\tau_s) \Bigg \} \nonumber \\ 
&&\hspace{.2in}+\lambda_l \Bigg \{ \frac{1}{1-\omega_l\tau_s} + \frac{1}{\omega_l} \int_0^\infty d\omega e^{-\omega/\omega_l}  \nonumber \\
&&\hspace{.5in} \times \left (\coth \left (\frac{\hbar \omega}{2k_BT}\right) -1\right) \sinh (\omega\tau_s) \Bigg \} . \label{eq:sta-soln-ohm} 
\een
It is clear from the above expression that the term involving $\lambda_l$ becomes singular at $\tau_s=1/\omega_l$.  There are additional singularities for larger values of $\tau_s$ as well, depending on the value of temperature.   This shows that the logarithmic dependence of $\tau_s$ with $ \Delta \bar E$, as expressed by Eq. (\ref{eq:tau_s-0}), is no longer a valid assumption for very large $\Delta \bar E$ even in the limit where $\lambda_h >> \lambda_l$.  

When the term  $\lambda_l/(1-\omega_l\tau_s)$ on the right hand side of Eq. (\ref{eq:sta-soln-ohm}) is dominant, the resulting solution for $\tau_s$ can be approximated as  $(1-\lambda_l/\Delta \bar E)/\omega_l$.   Thus, as a simple approximation, one can use the following expression for $\tau_s$: 
\be
\tau_s=\mbox{Min} \left\{ \frac{1}{\omega_h} \ln \left (\frac{\Delta \bar E}{\lambda_h}\right), \frac{1}{\omega_l} \left (1-\frac{\lambda_l}{\Delta \bar E}\right )\right\} , \label{eq:tau-s-approx}
\ee
where ``Min" represents the minimum of the two argument and it is assumed that $\Delta \bar E >0$.   

Employing $\tau_s$ given by Eq. (\ref{eq:tau-s-approx}) in Eq. (\ref{eq:kfg-5}) leads to a new generalized energy gap (GEG) law, which extends the original EG law, Eq. (\ref{eq:keg}), for high temperature and for the case where the solvent and low frequency modes  can be modeled by the Ohmic spectral density.   The expressions for $D(i\tau_s)$ and $G(\tau_s)$ used in Eq. (\ref{eq:keg}) are respectively the same as Eqs. (\ref{eq:d-tau}) and (\ref{eq:g-tau}) except that $\tau_s$ given by Eq. (\ref{eq:tau-s-approx}) is used instead of $\tau_s^0$.  
Although direct numerical integration involved in these terms is feasible, closed form expressions with reasonable accuracy for these would be useful and can be derived employing an approximation\cite{jang-jpcb106} for $\coth(x/2)\approx 1+2 e^{-x}+2e^{-2x}+(2/x) e^{-5x/2}$.  The resulting approximation for $D(i\tau_s)$ is as follows:
\ben
&&D(i\tau_s)\approx  \frac{\lambda_h \omega_h}{\hbar}\Bigg \{\exp(\omega_h\tau_s)\nonumber \\
&&\hspace{.6in}+ \left (\coth \left (\frac{\hbar\omega_h}{2k_BT}\right) -1\right )\cosh (\omega_h \tau_s )   \Bigg \}  \nonumber \\
&&+\frac{\lambda_l \omega_l}{\hbar}  \Bigg \{ \frac{1}{(1-\omega_l \tau_s)^2}  +\frac{1}{(1+\hbar\omega_l/(k_BT)-\omega_l \tau_s)^2}  \nonumber \\
&&\hspace{.6in} +\frac{1}{(1+2\hbar\omega_l/(k_BT)-\omega_l \tau_s)^2} \nonumber \\
&&\hspace{.6in}+\frac{1}{(1+\hbar\omega_l/(k_BT)+\omega_l \tau_s)^2} \nonumber \\
&&\hspace{.6in}+\frac{1}{(1+2\hbar\omega_l/(k_BT)+\omega_l \tau_s)^2} \nonumber \\
&&\hspace{.6in}+\frac{k_BT/(\hbar\omega_l)}{(1+5\hbar\omega_l/(2k_BT)-\omega_l \tau_s)} \nonumber \\
&&\hspace{.6in}  + \frac{k_BT/(\hbar\omega_l)}{(1+5\hbar\omega_l/(2k_BT)+\omega_l \tau_s)} \Bigg\} .  \label{eq:d-tau-geg}
\een
Similarly, $G(\tau_s)$ can also be approximated as
\ben
&&G(\tau_s)\approx  \frac{\lambda_h}{\hbar\omega_h}\Bigg\{\exp(\omega_h\tau_s) -1\nonumber \\
&&\hspace{.6in}+\left (\coth \left (\frac{\hbar\omega}{2k_BT}\right) -1\right )\left ( \cosh (\omega \tau_s ) -1\right)   \Bigg\} \nonumber \\
&&+\frac{\lambda_l}{\hbar\omega_l}  \Bigg \{ -\ln \left (1-\omega_l\tau_s\right ) \nonumber \\
&&\hspace{.6in}- \ln \left (1- \left (\frac{\omega_l\tau_s}{1+\hbar\omega_l/(k_BT)}\right)^2 \right)  \nonumber \\ 
&& \hspace{.6in}- \ln \left (1- \left (\frac{\omega_l\tau_s}{1+2\hbar\omega_l/(k_BT)}\right)^2 \right) \nonumber \\
&&\hspace{.6in}+\left (\frac{k_BT}{\hbar\omega_l} +\frac{5}{2}+\frac{k_BT \tau_s}{\hbar} \right) \nonumber \\
&&\hspace{.8in}\times  \ln \left (1+\frac{\omega_l\tau_s}{1+5\hbar\omega_l/(2k_BT)}\right ) \nonumber \\
&&\hspace{.6in}+\left (\frac{k_BT}{\hbar\omega_l} +\frac{5}{2}-\frac{k_BT \tau_s}{\hbar} \right)\nonumber \\
&&\hspace{.8in} \times \ln \left (1-\frac{\omega_l\tau_s}{1+5\hbar\omega_l/(2k_BT)}\right ) \Bigg\} .  \label{eq:g-tau-geg}
\een
For the above expressions, it is clear that both $D(i\tau_s)$ and $G(\tau_s)$ are finite for all values of $\tau_s$ given by Eq. (\ref{eq:tau-s-approx}).

\section{Numerical results and discussion}

This section compares both the EG rate expression, Eq. (\ref{eq:keg}), and the GEG rate expression, Eq. (\ref{eq:kfg-5}) with Eqs. (\ref{eq:tau-s-approx})-(\ref{eq:g-tau-geg}), against the exact evaluation of the FGR rate, Eq. (\ref{eq:kfg-2}), calculated by numerical fast Fourier transform.   In addition, both the semiclassical (SC) and stationary phase approximations with interpolation (SPI), which are appropriate for $|\Delta \bar E| \sim \lambda$,  are compared as well.  Detailed expressions for these SC and SPI approximations for the present case are provided by Eqs. (\ref{eq:ksc}) and (\ref{eq:kspi}), respectively, in Appendix B.  These methods and their conditions of validity are summarized in Table \ref{table1}.

\begin{table}
\caption{Equations and conditions of validity in addition to the stationary phase approximation for the methods tested. \label{table1}}
\begin{tabular}{c|c|c}
\hline
\hline
Rate & Expression & Conditions of validity \\
\hline
\multirow{2}*{Exact} & \multirow{2}*{Eq. (\ref{eq:kfg-2})} &  \\
&&\\
\hline
\multirow{2}*{EG} & \multirow{2}*{Eq. (\ref{eq:keg})}  & $\Delta \bar E >> \lambda_h >>\lambda_l$ \\
&& $\hbar\omega_h >> k_BT$\\
\hline
\multirow{2}*{GEG} & \multirow{2}*{Eq. (\ref{eq:kfg-5}) with  Eqs. (\ref{eq:tau-s-approx})-(\ref{eq:g-tau-geg})} & \multirow{2}*{$\Delta \bar E >>\lambda_h,\lambda_l$}  \\
& &  \\
\hline
\multirow{2}*{SC} & \multirow{2}*{Eq. (\ref{eq:ksc})} & \multirow{2}*{$\Delta \bar E \sim \lambda$}\\
&&\\
\hline
\multirow{2}*{SPI} & \multirow{2}*{Eq. (\ref{eq:kspi})} & \multirow{2}*{$\Delta \bar E < 2\lambda$} \\
&&\\
\hline
\hline
\end{tabular} 
\end{table}

\begin{table}
\caption{Values of parameters defining four different cases tested and example set of parameters for the case where $k_BT=200\ {\rm cm^{-1}}$.  In all the cases, $\delta \bar g_h^2=0.1$. \label{table2} }
\begin{tabular}{c|c|c|c|c|ccccc}
\hline
\hline
\multirow{2}*{Case} & \multirow{2}*{\makebox[.15in]{$\eta_l$}}  &  \multirow{2}*{\makebox[.15in]{$\frac{\omega_h}{\omega_l}$}} &\multirow{2}*{\makebox[.15in]{$\gamma$} }& \multirow{2}*{\makebox[.3in]{$\frac{k_BT}{\hbar\omega_l}$}} & \multicolumn{5}{c}{$k_BT=200\ {\rm cm^{-1}}$} \\ 
\cline{6-10} 
&&&&&\makebox[.3in]{$\hbar\omega_l$} &\makebox[.3in]{$\hbar\omega_h$}&\makebox[.3in]{$\lambda_l$} &\makebox[.3in]{$\lambda_h$} &$({\rm cm}^{-1})$\\
\hline
\multirow{2}*{I}  & \multirow{2}*{$1$} & \multirow{2}*{$5$} &\multirow{2}*{$\frac{1}{3}$}  & $1$  &200&1000& 200 & 100 \\
\cline{5-10}
 & && &$0.5$&400&2000&400 & 200\\
\hline
\multirow{2}*{II}  & \multirow{2}*{$2$} & \multirow{2}*{$5$}  & \multirow{2}*{$\frac{1}{5}$} &$1$ &200&1000&400 & 100\\
\cline{5-10}
  & & & & $0.5$&400&2000&800 & 200\\
\hline
\multirow{2}*{III}  &\multirow{2}*{$1$} &\multirow{2}*{$10$}  & \multirow{2}*{$\frac{1}{2}$} & $1$ &200&2000&200 & 200\\
\cline{5-10}
  &  &  & & $0.5$&400&4000&400 & 400\\
\hline
\multirow{2}*{IV} & \multirow{2}*{$2$} & \multirow{2}*{$10$}  & \multirow{2}*{$\frac{1}{3}$} &$1$ &200&2000&400 & 200 \\
\cline{5-10}
  &  &   & &$0.5$&400&4000&800 & 400\\
\hline
\hline
\end{tabular}
\end{table}

\begin{figure}
\includegraphics[width=3in]{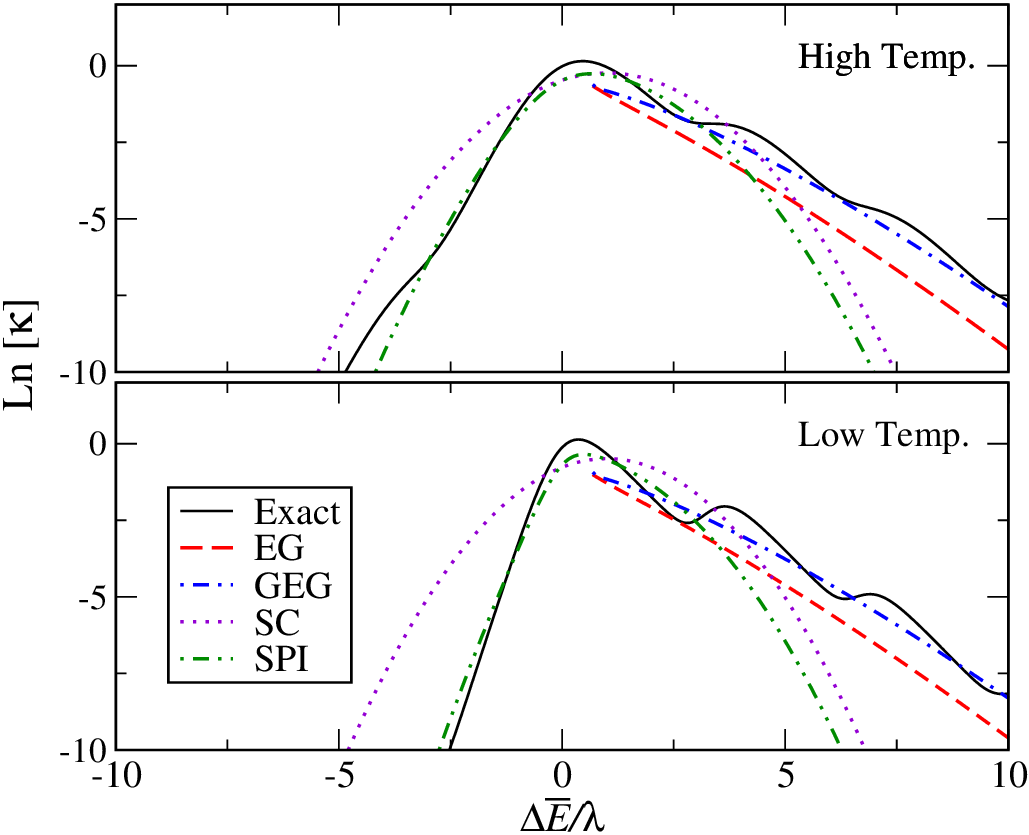}
\caption{Logarithms of dimensionless rate $\kappa$, which is obtained by multiplying an actual rate $k$ with $\hbar\sqrt{k_BT\lambda}/(\sqrt{\pi}J^2)$ for  Case I of Table \ref{table2}.   The upper panel (High Temp.) is for $k_BT=\hbar\omega_l$ and the lower panel (Low Temp.) is for $k_BT=\hbar\omega_l/2$.  Exact results are those obtained by exact numerical evaluation of the FGR rate expression.  EG, GEG, SC, and SPI respectively represents rates of the EG law, GEG law, semiclassical approximation, and the stationary phase approximation with interpolation. } 
\end{figure}

\begin{figure}
\includegraphics[width=3in]{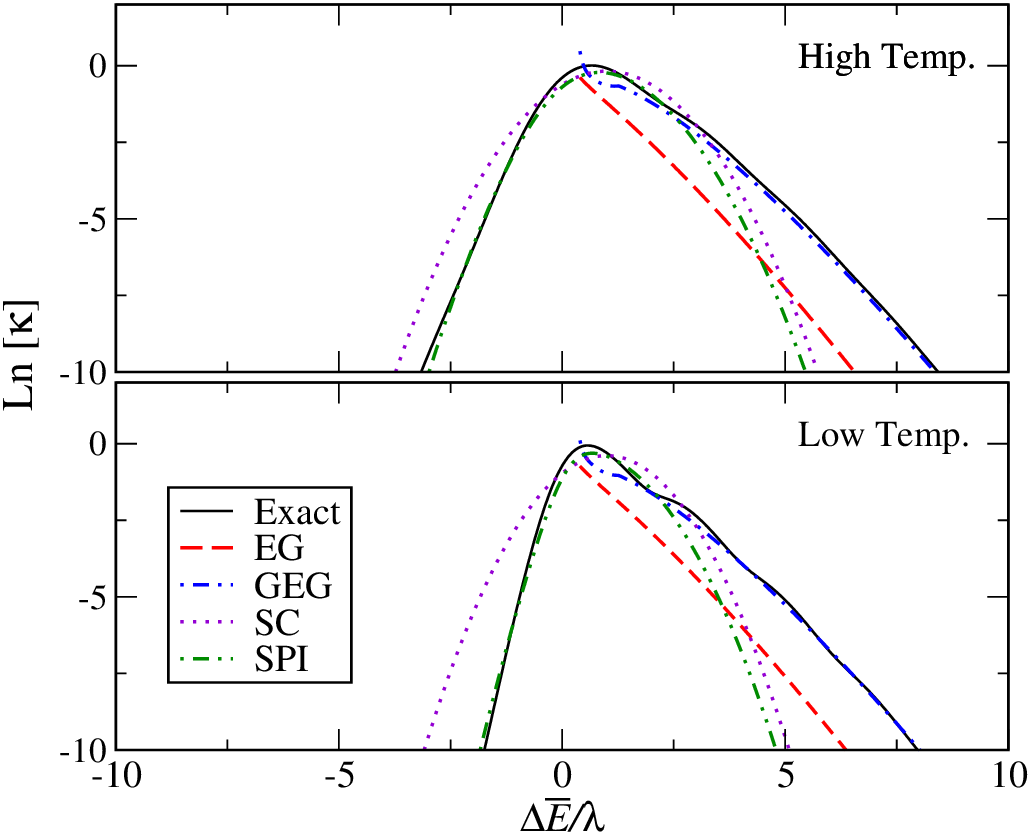}
\caption{Logarithms of dimensionless rate $\kappa$ for Case II of Table \ref{table2}. All other conventions are the same as Fig. 1.}
\end{figure}

The first case being tested (Case I) is for $\eta_l=1$ and $\omega_h/\omega_l=5$ while $\delta \bar g_h^2=\lambda_h/(\hbar\omega_h)=0.1$.  The value of $\gamma$ defined by Eq. (\ref{eq:gamma}) is $1/3$ for this case.  Two representative values of temperature, $k_BT/(\hbar\omega_l)=1$ and $0.5$, are shown.  Although the former is denoted as high temperature and the latter as low temperature, this distinction is relative to the value of $\omega_l$ and does not necessarily imply specific values of temperature. In other words,    the parameters chosen can represent a range of different situations depending on the actual value of $\omega_l$ or $k_BT$.   As an example, Table \ref{table2} provides actual values of parameters for the case where $k_BT=200\ {\rm cm^{-1}}$.   Figure 1 shows the results of calculation in terms of the logarithms of dimensionless rate (see the figure caption) with respect to the  energy gap relative to the reorganization energy, $\Delta \bar E/\lambda$, where $\lambda=0.1 \hbar\omega_h+\hbar\omega_l=1.5 \hbar\omega_l$ for the present case.   Both EG and GEG rate expressions were evaluated only for $\Delta \bar E \geq \hbar\omega_l$.  

The upper panel in Fig. 1 corresponds to the case with moderate quantum effects due to ${\mathcal J}_l(\omega)$.   This is the regime where the EG law is expected to work qualitatively but not all the conditions of the EG law are satisfied.  Indeed, both EG and GEG expressions reproduce nearly exponential dependence of the rate on the energy gap, which becomes apparent for $\Delta \bar E/\lambda >2$.   However, the former significantly underestimates whereas the latter agrees well with the exact numerical evaluation of the FGR rate.  The slight oscillatory pattern in the exact FGR result in this regime corresponds to the vibrational progression due to the high frequency mode and is not captured well by the stationary phase approximation.     

The lower panel in Fig. 1 corresponds to the case where the quantum effects due to ${\mathcal J}_l(\omega)$ are more enhanced.  Thus, there is significant discrepancy between the SC approximation and the FGR rate at all values of $\Delta \bar E/\lambda$ except at near $1$.  On the other, the performance of SPI remains  reliable for $\Delta \bar E/\lambda < 2$.  
Both EG and GEG reproduce the average qualitative behavior except for the oscillatory pattern, which is more pronounced than the high temperature case. As in the high temperature case, the former underestimates whereas the latter agrees well with the exact numerical evaluation of the FGR rate, although the performance of the former is slightly better than in the high temperature case.

Figure 2 shows results for Case II of Table \ref{table2},  for which the coupling to the low frequency bath spectral density, $\eta_l=2$, while others remain the same as those for Case I.    The reorganization energy $\lambda=0.1 \hbar\omega_h+2\hbar\omega_l=2.5 \hbar\omega_l$ for the present case. The results for the SC approximation in this case are closer to the FGR rate than those in Fig. 1 but are not as good as the SPI results, which show very good agreement with the FGR rates for $\Delta \bar E/\lambda <2$ in both high and low temperature cases.  As expected, the discrepancies between the EG rate and the FGR  rate are larger than those for Case I.  On the other hand, the GEG rate shows very good agreement with the FGR rate for $\Delta \bar E/\lambda >2$. The fact that the FGR results do not exhibit significant oscillatory pattern is consistent with the fact that the stationary phase approximation works well for this case.       
\begin{figure}
\includegraphics[width=3in]{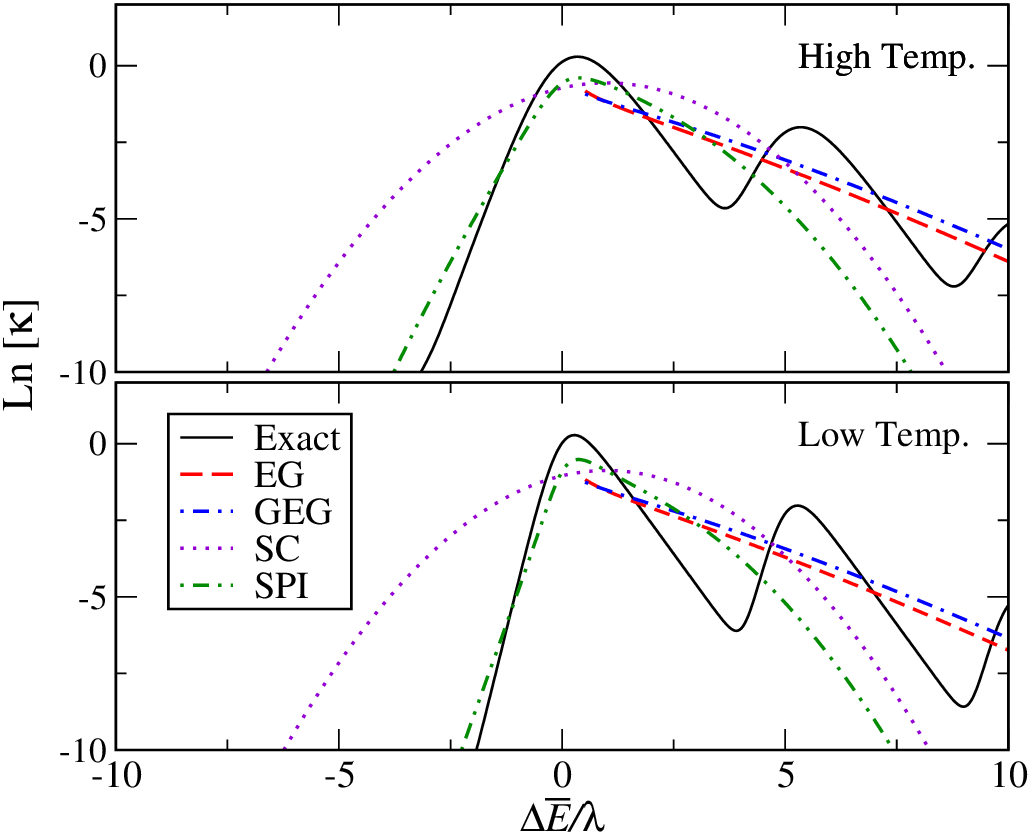}
\caption{Logarithms of dimensionless rate $\kappa$ for Case III of Table \ref{table2}.  All other conventions are the same as Fig. 1. }
\end{figure}

\begin{figure}
\includegraphics[width=3in]{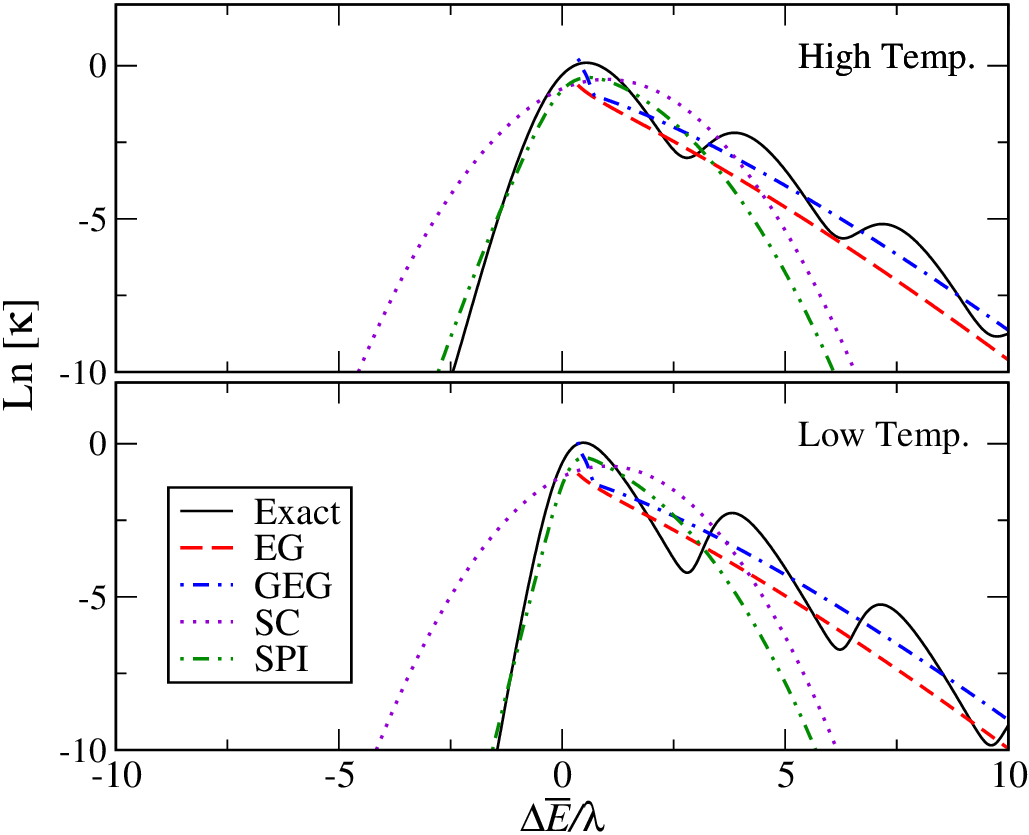}
\caption{Logarithms of dimensionless rate $\kappa$ for Case IV of Table \ref{table2}.  All other conventions are the same as Fig. 1.}
\end{figure}

Figures 3 and 4 show results for Cases III and IV of Table \ref{table2}, respectively.  Thus, the values of reorganization energy are $\lambda=0.1 \hbar\omega_h+\hbar\omega_l= 2\hbar\omega_l$ for Fig. 3 and $\lambda=0.1 \hbar\omega_h+2\hbar\omega_l= 3\hbar\omega_l$ for Fig. 4.  The larger difference between $\omega_h$ and $\omega_l$ in this case makes the EG rate and the GEG rate closer compared to those of Figs. 1 and 2.  However, the exact FGR results in this case exhibit stronger oscillatory pattern, which corresponds to pronounced vibrational sidebands in lineshapes that cannot be captured well by the stationary phase approximation.  Comparison of results between Figs. 3 and 4 shows that stronger value of $\eta_l$ dampen such oscillatory behavior, which on the other degrades the performance of the EG rate.

The numerical results shown in Figs. 1-4, although not exhaustive, offer important insights into the performance and limitations of the EG law. Overall, the EG rate, Eq. (\ref{eq:keg}), captures the gross qualitative trends of the dependence of the rate on the value of EG fairly well.   Although temperature dependences not represented in the EG rate can be seen, they do not appear to be significant either.  However, two quantitative deficiencies of the EG law can be identified clearly. 

First, for the cases where the stationary phase approximation is reasonable, as can be seen from the good performance of the GEG rate (Figs 1-2), the EG rate significantly underestimates the FGR rate.  This error can cause the nonadiabatic coupling constants determined from fitting experimental data by the EG rate to be much larger than its actual value. 

Second, for larger value of the high frequency vibrational modes (Figs. 3-4), where the assumption by EJ\cite{englman-mp18} is expected to be better justified as can be seen from relatively small differences between the EG and GEG rates, the stationary phase approximation becomes more inaccurate, resulting in substantial oscillatory features in the FGR rates.  Unless experimental results are classified according to the frequencies of the highest vibrational modes that are actively coupled to the electronic transitions, such oscillatory patterns may be difficult to be distinguished from scatters caused by experimental errors of unknown sources.  

The two issues of the EG rate addressed above suggest the difficulty of identifying the parameter regime where it serves as a reliable guidance for a broad range of experimental data.  The GEG rate and numerical tests provided in this section offer the possibility of resolving these issues to some extent.  First, the use of the GEG rate will enable more accurate assessment of the nonadiabatic coupling.  Although results were shown only for two representative values of temperature, the performance of the GEG rate does not degrade with temperature when tested.  This is because it incorporates all contributions of thermally excited states of the bath vibrational modes.  Second, Figs. 3-4 show the possibility of experimentally observing oscillatory patterns in the dependence of nonradiative decay rates on the value of EG, through more judicious classification of experimental data aided by computational information.  Once this is achieved, the interval of oscillation can provide a direct evidence for the frequency of vibrational modes coupled to the electronic transition.

In addition to the issues concerning the EG rate as addressed above, outcomes of numerical tests in this section offer the possibility and potential utility of more general rate expression based on the stationary phase approximation. The overall results show that the SPI approximation works best for $\Delta \bar E/\lambda < 2$ and the GEG expression works best for  $\Delta \bar E/\lambda > 2$.  Thus, it is reasonable to consider the following empirical global interpolation expression:
\be
k_{GI}=\frac{k_{SPI}}{1+e^{\Delta \bar E/\lambda -2}} + \frac{k_{GEG}}{1+e^{-\Delta \bar E/\lambda +2}} .  \label{eq:k-gi}
\ee
While this expression is entirely based on empirical observation and does not have clear physical basis, test for various parameters that have been conducted suggest that this serves as a practically useful expression for describing and modeling the dependence of the rate on $\Delta \bar E$ without any restriction for a broad range of temperature, as shown before for the case of ET rate.\cite{jang-jpcb110}   As yet, with more careful theoretical consideration and computational test, there seems to be a good possibility to come up with similar or even better expression with better physical basis.  Considering recent interest and the importance of quantitatively modeling of nonradiative rates for wide range of EG values, for example, for ET,\cite{jones-jpca124} exciton transfer,\cite{ishizaki-jpcb125} singlet fission,\cite{yost-nc6} and reverse intersystem crossing,\cite{kim-jctc16} the above expression or more satisfactory version that may be developed can certainly help more quantitative modeling of such rate processes. \vspace{.1in}\\

\section{Conclusion}
The present work has clarified the nature of approximation involved in the EG rate derived by EJ\cite{englman-mp18} more than fifty years ago.  Thus, it was demonstrated in detail that the EG rate, Eq. (\ref{eq:keg}), is a specific kind of stationary phase approximation for FGR that becomes valid when the transition rate is dominated by ground vibrational state of the highest frequency mode coupled to the electronic transition.  Then, a simple generalization, GEG, which is applicable for the case where the bath spectral density consists of a narrow distribution of high frequency modes and an Ohmic spectral density was derived.   Test of EG and GEG rates for representative models confirms that the latter significantly improves the former, which can serve as better expression for modeling of experimental data. The numerical data also confirmed the regime where both are applicable is  relatively large value of $\Delta \bar E$ compared to the reorganization energy $\lambda$, typically for $\Delta \bar E/\lambda >2$.   This region complements the region where the SPI expression\cite{jang-jpcb110} works well, $\Delta \bar E/\lambda <2$.  Based on this observation, a simple interpolation formula,  $k_{GI}$ given by Eq. (\ref{eq:k-gi}), was suggested.  Further test and improvement of this expression will help understand and model more general cases of the energy dependence of nonradiative processes.  

It is also important to note that the assumption of weak coupling is not always consistent with the stationary phase approximation as can be seen from the comparison with direct numerical evaluation of the FGR rate.  So far, it is difficult to identify comprehensive experimental data confirming the oscillatory pattern of the EG dependence of the rate, to the best of my knowledge, but there is no reason why it cannot be realized experimentally and then utilized for controlling non-radiative decay rates as has been demonstrated recently.\cite{kim-jacs-au-1}

The model and theoretical analysis presented in this work are more general than those used by EJ.\cite{englman-mp18} The specific case where the low frequency bath is represented by the Ohmic spectral density with exponential cutoff is also widely applicable.  However, there remain important aspects and issues that need further theoretical and computational investigations.  The first important issue is to account for the effects of more structured bath spectral densities, {\it e.g.}, as can be found in photosynthetic light harvesting complexes.\cite{jang-rmp90}  Detailed test of the EG or GEG rates (to the extent that approximation with an Ohmic bath spectral density is feasible) for such cases and development of further generalization are necessary, which will be the topic of future research. The second issue is the contribution of more complicating factors such as non-Condon effect\cite{niu-jpca114,sun-jcp144} and Duschinsky rotation.\cite{ianconescu-jpca108,peng-jcp126,niu-jpca114,kim-jctc16}  While direct evaluation of the FGR rate for these cases is feasible, a simple generalization of the EG law for these cases, if available, will help more quantitative modeling of actual experimental data.

\acknowledgments
The author thanks Justin Caram for bringing his attention to the EG law, for sharing unpublished manuscript, and for helpful comments.   This research was supported by the US Department of Energy, Office of Sciences, Office of Basic Energy Sciences (DE-SC0021413).

\begin{center}
{\bf  Author Declarations} \\
\end{center}
The author has no conflicts to disclose. 

\begin{center}
{\bf  Data Availability Statements} \\
\end{center}
Most data that support the findings of this article are contained in this article.  Additional data are available from the corresponding author upon reasonable request.

\appendix 
\section{First order correction for $\tau_s^0$ for general case}
This section provides the first order correction of $\tau_s^0$. 
Let us introduce $\delta \tau_s$ such that 
\be
\tau_s=\tau_s^0+\delta \tau_s ,
\ee 
and make the following linear approximations, 
\ben
\exp (\omega \tau_s) &\approx & \exp(\omega \tau_s^0)(1+\omega\delta \tau_s)  ,  \\
\sinh (\omega \tau_s) &\approx & \sinh (\omega \tau_s^0)+\cosh (\omega \tau_s^0) \omega \delta \tau_s .
\een
Employing these expressions for both $\omega_h$ component and the low frequency bath spectral density term, Eq. (\ref{eq:sta-soln-1}) can be expressed as follows:
\ben
\Delta \bar E &\approx&    \Delta \bar E(1+\omega_h\delta \tau_s)\nonumber \\
& +& \lambda_h \left (\coth \left (\frac{\hbar \omega_h}{2k_BT}\right )-1\right) \nonumber\\
&&\hspace{.3in} \times  \left (\sinh (\omega_h \tau_s^0) +\cosh (\omega_h \tau_s^0)\omega_h \delta \tau_s \right )\nonumber \\
&+&\frac{1}{\pi} \int_0^{\infty} d\omega \frac{{\mathcal J}_l(\omega)}{\omega} \Bigg \{ \exp (\omega\tau_s^0) (1+\omega\delta \tau_s) \nonumber \\
&&\hspace{.2in}+ \left (\coth \left (\frac{\hbar \omega}{2k_BT}\right) -1\right) \nonumber \\
&&\hspace{.3in}\times  \left (\sinh (\omega \tau_s^0) +\cosh (\omega \tau_s^0)\omega \delta \tau_s \right )\Bigg \} \ ,
  \label{eq:sta-soln-2} 
\een    
where the definition for $\tau_s^0$, Eq. (\ref{eq:tau_s-0}), has been used.  Collecting all the terms proportional to $\delta \tau_s$ on the lefthand side and dividing the equation with $-\omega_h$, we obtain
\ben 
&&\Bigg [\Delta \bar E +\lambda_h \left (\coth \left (\frac{\hbar \omega_h}{2k_BT}\right) -1\right)\cosh (\omega_h \tau_s^0)\nonumber \\
&&\hspace{.2in}+\frac{1}{\pi\omega_h} \int_0^{\infty} d\omega {\mathcal J}_l(\omega) \Bigg \{ \exp (\omega\tau_s^0)  \nonumber \\
&&\hspace{.4in}+ \left (\coth \left (\frac{\hbar \omega}{2k_BT}\right) -1\right) \cosh (\omega \tau_s^0)  \Bigg \} \Bigg ]\delta \tau_s \nonumber \\
&&\approx- \frac{\lambda_h}{\omega_h}\left (\coth \left (\frac{\hbar \omega_h}{2k_BT}\right) -1\right)\sinh (\omega_h \tau_s^0)\nonumber \\
&&\hspace{.2in}-\frac{1}{\pi\omega_h} \int_0^{\infty} d\omega \frac{{\mathcal J}_l(\omega) }{\omega}\Bigg \{ \exp (\omega\tau_s^0)  \nonumber \\
&&\hspace{.4in}+ \left (\coth \left (\frac{\hbar \omega}{2k_BT}\right) -1\right) \sinh (\omega \tau_s^0)  \Bigg \}  .
\een
Employing the definition of $\tau_s^0$ explicitly and rearranging terms so that the order of $\lambda_h/\bar E$ appear clearly, the above expression can be shown to be equivalent to the following expression:
\ben 
&& \Bigg [ \frac{1}{2}\left (1+\coth \left (\frac{\hbar\omega_h}{2k_BT}\right) \right )\nonumber \\
&&\hspace{.2in}+\frac{1}{2} \left (\frac{\lambda_h}{\Delta \bar E}\right)^2 \left (\coth \left (\frac{\hbar\omega_h}{2k_BT}\right) -1 \right ) \nonumber \\
&&\hspace{.2in}+\frac{1}{\Delta \bar E \pi \omega_h}\int_0^\infty d\omega {\mathcal J}_l(\omega)  \Bigg \{ \coth \left (\frac{\hbar\omega}{2k_BT}\right)  \nonumber \\ 
&&\hspace{.6in}\times \frac{1}{2} \left ( \left ( \frac{\Delta \bar E}{\lambda_h}\right)^{\omega/\omega_h} -\left (\frac{\lambda_h}{\Delta\bar E}\right)^{\omega/\omega_h} \right )  \nonumber \\
&&\hspace{.4in}+\frac{1}{2} \left ( \left ( \frac{\Delta \bar E}{\lambda_h}\right)^{\omega/\omega_h} +\left (\frac{\lambda_h}{\Delta\bar E}\right)^{\omega/\omega_h} \right ) \Bigg \}\Bigg ]\delta \tau_s \nonumber \\
&&\approx -\frac{1}{\omega_h} \left (\coth \left (\frac{\hbar\omega_h}{2k_BT}\right)-1\right)\frac{1}{2} \left (1-\left (\frac{\lambda_h}{\Delta \bar E}\right )^2\right ) \nonumber \\
&&-\frac{1}{\Delta \bar E\pi \omega_h} \int_0^\infty d\omega \frac{{\mathcal J}_l(\omega)}{\omega}  \Bigg \{ \coth \left (\frac{\hbar\omega}{2k_BT}\right)  \nonumber \\ 
&&\hspace{.6in}\times \frac{1}{2} \left ( \left ( \frac{\Delta \bar E}{\lambda_h}\right)^{\omega/\omega_h} -\left (\frac{\lambda_h}{\Delta\bar E}\right)^{\omega/\omega_h} \right )  \nonumber \\
&&\hspace{.4in}+\frac{1}{2} \left ( \left ( \frac{\Delta \bar E}{\lambda_h}\right)^{\omega/\omega_h} +\left (\frac{\lambda_h}{\Delta\bar E}\right)^{\omega/\omega_h} \right ) \Bigg \} . 
\een
Dividing the above expression with the term inside $[ \cdots ]$, which is multiplied to $\delta \tau_s$ on the left side, leads to a general expression for $\delta \tau_s$ that is independent of the type of ${\mathcal J}_l(\omega)$ (to the extent that integrations involving this spectral density are well defined) and is valid as long as the deviation $\delta \tau_s$ is much smaller than $\tau_s^0$.  Application of the resulting expression $\tau_s=\tau_s^0+\delta \tau_s$ in Eq. (\ref{eq:kfg-eg}) then results in another kind of GEG that is applicable when small corrections for the original EG law, Eq. (\ref{eq:keg}), are valid.    This expression will be useful for the case where ${\mathcal J}_l(\omega)$ is directly obtained from simulation and cannot be modeled as a simple Ohmic form. 
\section{Stationary phase approximations for $|\Delta \bar E| \sim \lambda$}
The known stationary phase approximations for the ET reaction can easily be converted for the present case by using the appropriate definition of the spectral density given here and using the fact that $\Delta G=-\Delta \bar E$.   First, for the given spectral density, the three quantum reorganization energies\cite{jang-jpcb110} that appear in the stationary phase approximations near  $|\Delta \bar E| \sim \lambda$ can be expressed as
\ben
\lambda_{q,c} &\equiv& \frac{\hbar}{2\pi k_BT} \int_0^\infty {\mathcal J} (\omega) \coth \left (\frac{\hbar \omega}{2k_B T}\right ) \nonumber \\
&=&\lambda_h\frac{ \hbar\omega_h}{2k_BT} \coth \left (\frac{\hbar\omega_h}{2k_BT}\right )\nonumber \\
&&+\frac{\hbar}{2\pi k_BT} \int_0^\infty {\mathcal J}_l(\omega) \coth \left (\frac{\hbar \omega}{2k_B T}\right ) , \\
\lambda_{q,s} &\equiv& \frac{\hbar}{2\pi k_BT} \int_0^\infty {\mathcal J} (\omega) \frac{1}{\sinh \left (\hbar \omega/(2k_B T)\right )} \nonumber \\
&=&\lambda_h\frac{ \hbar\omega_h}{2k_BT} \frac{1}{\sinh \left (\hbar\omega_h/(2k_BT)\right )}\nonumber \\
&&+\frac{\hbar}{2\pi k_BT} \int_0^\infty {\mathcal J}_l(\omega)\frac{1}{\sinh \left (\hbar \omega/(2k_B T)\right )}   , \\
\lambda_{q,t} &\equiv& \frac{4k_BT}{2\pi \hbar} \int_0^\infty \frac{{\mathcal J} (\omega)}{\omega^2} \tanh \left (\frac{\hbar \omega}{4k_B T}\right ) \nonumber \\
&=&\lambda_h\frac{ 4k_BT}{\hbar\omega_h} \tanh \left (\frac{\hbar\omega_h}{4k_BT}\right )\nonumber \\
&&+\frac{4k_BT}{\pi\hbar} \int_0^\infty \frac{{\mathcal J}_l(\omega)}{\omega^2} \tanh \left (\frac{\hbar \omega}{4k_B T}\right ) .
\een
The SC approximation\cite{hopfield-pnas71,jang-jpcb110} for the rate is then given by 
\be
k_{sc}=\frac{J^2}{\hbar^2} \left (\frac{\pi\hbar^2}{k_BT\lambda_{q,c}}\right)^{1/2} \exp\left \{-\frac{(\Delta \bar E-\lambda)^2}{4k_BT\lambda_{q,c}}\right\} . \label{eq:ksc}
\ee
On the other hand, the SPI approximation\cite{jang-jpcb110} for the present case is given by 
\be
k_{SPI}=\frac{J^2}{\hbar^2} \left ( \frac{\pi\hbar^2 W}{k_BT \lambda_{q,in} }\right)^{1/2}  \exp\left \{-\frac{(\Delta \bar E-\lambda)^2}{4k_BT\lambda_{q,c}}\right\} ,  \label{eq:kspi}
\ee
where 
\ben
&&\lambda_{q,in}=(\lambda_{q,c}-\lambda_{q,s})\left (\frac{\Delta \bar E}{\lambda}\right)^2+\lambda_{q,s} , \\
&&W=\frac{1}{1+\alpha e^{-\gamma \Delta \bar E/\lambda}} ,
\een 
with
\ben 
&&\alpha=e^{(\lambda_{q,c}\lambda_{q,t}-\lambda^2)/(2k_BT\lambda_{q,c})}-1 , \\
&&\gamma=\ln \left [\frac{e^{2\lambda (\lambda_{q,c}-\lambda)/(k_BT\lambda_{q,c})}-1}{e^{(\lambda_{q,c}\lambda_{q,t}-\lambda^2)/(2k_BT\lambda_{q,c})}-1}\right ] .
\een

\end{document}